\documentstyle[12pt]{article}
\begin{document}
\baselineskip=24pt
\setcounter{page}{1}
\parskip=0pt plus2pt
\def\be{\begin{equation}}
\def\en{\end{equation}}
\def\bea{\begin{eqnarray}}
\def\ena{\end{eqnarray}}
\def\de{\delta}
\def\De{\Delta}
\def\Dev{\Delta_v}
\def\r{\rho}
\def\al{\alpha}
\def\f{\frac}
\def\P{\Phi}
\def\w{\wedge}
\def\rc{\widehat{\rho}}
\def\no{\noindent}
\def\cf{{\cal F}}
\def\cp{{\cal P}}
\def\cv{{\cal v}}
\def\lb{\left[}
\def\rb{\right]}
\begin{titlepage}
\begin{center}
{\Large\bf Stability of Amorphous Structures with Voids}

\vspace*{.5cm}
\end{center}
\begin{center}
{\it Charanbir kaur and Shankar P. Das }\\
{\it School of Physical Sciences,\\
  Jawaharlal Nehru University,\\
  New Delhi 110067, India. }

\end{center}
\begin{center}
{\large{Abstract}}\\
\end{center}
We incorporate the role of free volume in the density function of the
amorphous
structure and study its effects on the stability of such structures. The
Density Functional Theory is used to explore this ``Free Volume Model''
of the supercooled  structures.
The Free energy minimization is carried out using the void concentration
as a variational parameter. A critical value of this concentration 
exists corresponding to the Free energy minima of the amorphous structure.
An increase in the stability  is observed due to the inclusion of voids in
the density structure.
This study is conducted for both the weakly and highly localized amorphous
structures. The free volume concentration shows a power law
decrease with density for the weakly localized states and a linear decrease
for the highly localized amorphous structures. 

\vspace*{2cm}
\noindent
PACS number(s) : 64.70Pf, 64.60 Cn, 64.70 Dv
\end{titlepage}

\section*{I Introduction}
The transformation of a continuum ergodic liquid into an amorphous glassy
state with solid-like behaviour has been an area of much current research
interest. The ambiguities concerning the glassy or the supercooled
state have led to the development of a large number of theoretical
approaches \cite{CT,MCT,SW,GP} that are used to study the basic
characteristics of a system undergoing such a transition. These techniques
explore both the dynamic and the thermodynamic properties of the supercooled
states. A liquid is supercooled while avoiding crystallization such that its
density structure loses the homogeneity of a liquid state and attains a
heterogeneous density
profile. From a thermodynamic point of view, the supercooled state represents
a metastable state relative to the crystal and is characterized by an
inhomogeneous and aperiodic density structure. This aperiodic crystal
picture of the supercooled state is considered here to study its stability
from a thermodynamic approach.
In this work, we intend to explore the ``Free Volume'' picture of the
supercooled liquid \cite{CT,CG} by studying the Free energy landscape of an
amorphous lattice characterized by a finite fraction of vacancies.

Fox and Flory \cite{FOX} first postulated that the liquid-glass transition
resulted from the decrease of ``free volume'' of the amorphous phase
below some critical value. Subsequently, this concept was further elaborated
by Cohen $\&$ Turnbull \cite{CT} and later by Cohen $\&$ Grest \cite{CG} who
formally developed the theoretical aspects using the percolation theory.
The basic phenomenology behind the Free Volume models is that transport of
particles occurs by their flow into voids of a size greater than a critical
value formed by the redistribution of some ``free volume''. The latter is
loosely defined as some surplus volume that is not taken up by the particles.
As a consequence of this theory, the relation between the free volume
concentration and the transport properties of the supercooled state like the
viscosity was also investigated. Doolittle \cite{DO} found  empirically that
the fluidity (inverse of viscosity) of many simple hydrocarbon liquids could
be represented in terms of free volume as, 
\be
\label{DE}
\phi=\phi_o exp \left( -\f{c~ v_o}{v_f} \right)
\en
where $\phi$ denotes the fluidity and $c$ is a constant of order unity. Here
$v_f$ denotes the average free volume per molecule, $v_f={\bar v}-v_o$,
${\bar v}$ being the average volume per molecule and $v_o$ is the van der
Waals volume of the molecule. Williams  $et al.$ showed that this
result is valid for a large number of glass formers. They  proposed that a
linear relation exists between the free volume $v_f$ and the distance from
transition on the temperature scale $T-T_g$. This, through eqn.(\ref{DE}),
in turn gives rise to the typical Vogal-Fulcher divergence of viscosity.

The Density Functional Theory (DFT) is an appropriate liquid based theory to
undertake this study. This theory is based upon the simple ideology of
identifying a solid as a strongly nonuniform liquid. This notion of
describing the properties of a highly non-uniform phase in terms of the
fairly well-developed liquid state theory was first put forward by Kirkwood
and Monroe \cite{KM}. Later on,
Ramakrishnan and Yussouf (RY) \cite{RY} formulated it in terms of direct
correlation function and successfully applied this concept to liquid-solid
coexistence problem by expressing the properties of the
highly organized solid( the non-uniform phase) as a perturbation of those
of the coexisting liquid. The mathematical formulation of this simple
concept was further elucidated by Haymet and Oxtoby \cite{OX} and a fair
amount of success has since been achieved in solving problems related to
simple inhomogeneous density structures. This theory has been lately applied
to  study the static aspects of supercooled amorphous structures.
These studies were mainly initiated by Singh $et.al.$ \cite{SW} and were
followed by similar studies using different density functional treatments
\cite{BC,HL}. Two types of Free energy minima were located for such systems.
The first
corresponding to highly localized amorphous structures that
were identified as the ``hard sphere glass'' state \cite{SW,BC,HL}. A different
class of
minima were detected \cite{SPD1} in the Free energy landscape of the same
random structure,
that represented {\em weakly localized} heterogeneous structures. These
structures were shown to be more realistically associated with the structure
of supercooled states seen in computer simulation studies.
The metastable character of
the different type of minima in the free energy  landscape by modeling
``free volume'' in the
density function characteristic of these structures is explored here. This has the effect
of increased heterogeneous character of the underlying amorphous structure.
We compute the Free energy
functional using the RY expansion around  a uniform state.

The motivation of the present work is 
to formulate the concept of free volumes in a simple density
functional formalism and use its variational nature to investigate whether
there
is any optimum value of the free volumes associated with the metastable
phase, giving a free energy minima.
In the model presented here, the free volumes in the system are constructed
as the vacancies at the prescribed lattice sites of the Bernal's random
structure \cite{BER}. Such a modeling leads to a scenario of increased
amount of local unoccupied volume in the system and that is how we define
``free volume'' in this work. The major requirement of such theories is the
formulation of the test density function which is relevant to the type of
structure whose stability is being considered. The variational principle is
then invoked to determine the density structure corresponding to the minimum
of the Free energy functional. P. Tarazona \cite{PT} first proposed to
consider the ensemble averaged  density distribution as a summation over
Gaussian profiles centered over the lattice sites of the given inhomogeneous
structure. Since then, such a formulation has been followed  almost
uniformly in all the related Density Functional studies.
In this approach, the density is parameterized in terms of the mass
localization parameter $\al$ which is proportional to the width of the
Gaussian profiles. This parameter has the effect of quantifying the
motion of particles in the system. The $\al \rightarrow 0$ limit depicts
Gaussian profiles of infinite width and thus the corresponding phase
represents the homogeneous liquid state. Increasing values of $\al$ represent
increasingly localized structures and thus referring to greater inhomogeneity
in the system. The $\al$ corresponding to the Free energy minima finally
determines the phase which the system prefers thermodynamically.
To include vacancies in the system, the parameter space of the density function
is increased. The homogeneous presence of a specific
concentration of vacancies in the system leads to a uniform decrease in the
number of occupied sites in each co-ordination shell around any arbitrary
position in the system. In an ensemble-averaged picture this amounts to
expressing the density function as,
\be
\label{rho}
\r(\vec r)=\cp (\f{\al}{\pi})^{\f{3}{2}} \sum_{i=1}^{N_L} e^{-\al |\vec r -
\vec  R_i|^2)}
\en
\noindent where $\cp$ actually serves as the fraction of sites present in
any $i^{th}$ shell around the position $\vec r$ in the system, such that
$\Dev=1-\cp$ represents the vacancy concentration in the system. We assume the
vacancies to be distributed homogeneously throughout the system and hence
this fraction is the same for all the shells around any arbitrary position
in the system. This approximation is followed from Jones and Mohanty
\cite{UM} who considered defects in a perfect crystal structure.
Here $N_L$ is the number of lattice sites in volume V out of which $N$
sites are occupied by the particles. The average particle density is thus,
$\r_o=\cp~\r_b$ where $\r_b$ represents the average lattice density.
Thus the Free energy minimization is carried out with respect to $\al$ and
$\cp$. These two parameters elaborate on two different structural aspects of
the stable phase. We mainly explore the existence of a critical free volume
concentration $\cp$ and its variation with density for the two different
class of minima seen for amorphous structures.

The paper is organized in the following manner : In section II we give the
details related to the calculation of the Free energy functional formulated
so as to include the effects of free volume.
In section III we evaluate the effects on the stability of the amorphous
structures by the induction of vacancies in the density structure. Here we
present the analysis conducted for amorphous system in both the extremely
localized and the weakly localized regimes. The implications of these
results are further discussed in the section IV and relevance with the
dynamical studies is quantitavely explored.

  \section*{II The Model Studied}
  The Helmholtz Free energy can be expressed as a sum of the purely
  entropic
  contribution given by the ideal gas term and the contribution due to the
  interactions in the system identified as the excess term in the general
  terminology.
  \be
  \label{free}
  \cf [\r(\vec r)]=\int d \vec r \r(\vec r) (ln(\w^3
  \r(\vec r))-1) - \P[\r(\vec r)]
  \en
  Here $\r(\vec r)$ is the ensemble averaged local density
  distribution and $\P$ represents the excess part of the Free energy.
  $\P[\r(\vec r)]$ is expressed as a perturbation around a liquid of
  density
  $\r_o$. The functional Taylor series expansion about the homogenous
  density distribution $\r_o$ is given as \cite{RY},
  \bea
  \P[\r(\vec r)]-\P(\r_o)&=& \int d\vec r_1 \de \r(\vec r_1) \lb-\beta
  \mu_o
  + ln(\w^3 \r_o) \rb \\
  \nonumber
  & &+\f{1}{2} \int d \vec r_1 \int d \vec r_2 c(\vec r_1
  -\vec r_2;\r_o) \de \r(\vec r_1 ) \de \r(\vec r_2)
  \ena
where $\mu_o$ is the chemical potential of the uniform phase  and $c(r;\r_o)$
is the direct correlation function of the homogeneous state of density $\r_o$.
  In the present work we consider the perturbation expansion of the
  excess Free energy around the uniform system of the same density as the
  average density of the corresponding inhomogeneous structure, i.e.,
  \be
  \r_o=\f{1}{V}\int d \vec r \r(\vec r)
  \en
  The Free energy expression in this case reduces to the following form,
  \bea
  \label{fre2}
  \cf[\r(\vec r)]-\cf(\r_o) &=& \int d \vec r \r(\vec r)~
  ln \left( \f{\r(\vec r)}{\r_o} \right) \\
  \nonumber
  & &-\f{1}{2}\int d \vec r_1 \int d \vec r_2 \de \r(\vec r_1) \de \r(\vec
  r_2)
  c(\vec r_1-\vec r_2;\r_o).
  \ena
\no Here $\de \r(\vec r)$ is the deviation of the inhomogeneous state density
$\r(\vec r)$ from the average density $\r_o$.
The minimization of this free energy functional then determines the density
function corresponding to the stable ( or metastable) state of the given
system. The first term is the purely entropic contribution to the Free energy
and the interactions in the system are responsible for the second term.

This functional is then minimized in a constrained
manner by the specification of the underlying lattice as the input.
This information is carried by the site-site correlation function $g(R)$
of the particular structure considered. Bernal's random structure \cite{BER}
is considered as a good approximation for the static structure of
undercooled liquid state. We follow the earlier approach of
supplying the $g(R)$ of the Bernal structure determined from the Bennett's
algorithm \cite{B} and approximate \cite{BC},
\begin{equation}
\label{gr}
g(\vec R) = g_{B}[ R(\frac{\eta}{\eta_o})^{\frac{1}{3}}]
\end{equation}
\noindent where $ \eta $ denotes the average packing fraction.
Here $ \eta_o $ is used as
a scaling parameter for the structure, \cite{BC,HL} such that at
$ \eta =\eta_o $ the structure corresponds to Bernal packing, so that
different values of $\eta_o$ relate to different structures.
Now we elaborate the computational aspects of reducing these expressions
for the specific case of a random structure.

With the density function expressed in terms of defects, i.e. eqn.(\ref{rho}),
the Ideal gas part of the Free energy per particle can in general be expressed
as,
\be
\label{EX}
\De {f}_{id}[\rho]= \cp \int d\vec r \phi(\vec r) \lb ln \left( \f{\cp}{\r_o}
\int d \vec R \phi( \vec r - \vec R) \left( \de (\vec R) + \rho_o
g( \vec R) \right) \right) \rb
\en
for an amorphous structure described by the radial distribution function $g(R)$
referred above. This is obtained by the use of eqn.(\ref{rho}) in the ideal
gas term of eqn.(\ref{fre2}). The random structure is marked by a vacancy
concentration of $\Dev$.
In the above expression, the function $\phi(\vec r-\vec R)$ represents a Gaussian
function centered at a position $\vec R$, i.e.,
$\phi(\vec r-\vec R)=(\f{\al}{\pi})^{\f{3}{2}} e^{-\al |\vec r - \vec R|^2)}$.
The density structure of the system in a highly localized state is represented
by sharply peaked Gaussian functions centered around the lattice sites. This
asymptotic limit corresponds to large $\al$ and the ideal gas part of free
energy (eqn.(\ref{EX})) reduces to,
\be
\label{app}
\De f_{id}[\rho] \approx -\lb \frac{3}{2} + ln~ \r_o - ln~ \cp +
\f{3}{2}~ln \left( \frac{\pi}{\al} \right) \rb.
\en
This is obtained by replacing the summation inside the logarithmic term
by just the contribution from the nearest site. This reduction is although
not valid for consideration of the weakly localized state of the amorphous
structure where the $\al$ values are small and the Gaussian functions wide
enough to cause considerable overlap over different sites such that the
above reduction can no longer be justified.
Thus in the range of small $\al$ we numerically evaluate the equation
(\ref{EX}) as such and the equation (\ref{app}) in the large $\al$ regions.
In our earlier work \cite{SPD1} we have shown that these two expressions
start merging for typically  $\al>40$ within $1\%$.
Similarly we formulate the evaluation of the excess term
per particle as,
\bea
\label{int}
\De f_{ex}&=&\r_o \int d \vec r c(r;\r_o) -\f{\cp}{2}\int d \vec r_1 \int d
\vec r_2 \phi(\vec r_1) \phi(\vec r_2) c(\vec r_1-\vec r_2;\r_o)\\
\nonumber
& & -\f{\cp}{2} \int d \vec R \int d \vec r_1 \int d \vec r_2
\phi(\vec r_1) \phi( \vec r_2 - \vec R) \left( \de (\vec R) + \rho_o
g( \vec R) \right) c(\vec r_1-\vec r_2;\r_o)
\ena
Here the direct correlation function, $c(r)$, carries the liquid state
interaction information. For the description of $c(r)$ we have used the
Percus-Yevick (PY) form with Verlet-Weiss correction as obtained by
Henderson and Grundke \cite{VW,HG}. This concludes the formulation of the
free energy functional used to measure the effects of free volume on the
stability of amorphous systems.

\section*{III Stability of the amorphous structure}
We now elaborate on the methodology used to investigate the stability of
the amorphous structure characterized by a finite vacancy fraction $\Dev$.
The free-energy functional is numerically evaluated using the expressions
(\ref{EX}) to (\ref{int}) at a given average density $\r_o$. The parameters
$\al$ and $\cp$  are varied simultaneously and two minima are detected in
the two dimensional ($\al$, $\cp$) free energy landscape. One corresponds
to the weakly localized state of the amorphous structure since it is found
corresponding to low $\al$ values. The other corresponds to very highly
localized structures characterized by large $\al$ values.
We illustrate the minimum on the Free energy landscape
with respect to the variational parameter $\cp$ in the Figure 1. The plotted
values correspond to the minima in the large $\al$ space.
The minimum value of the Free energy is observed at $\cp=0.951 $ at density
$\r_o=1.12$, as indicated by the arrow.
This figure  depicts a considerable increase in the stability of the
structure by the induction of free volumes in the structure. The $\De f$
values decreases from $1.64 (\cp=1)$ to $-0.44 $ due to the presence of
free volume in the system. As an inset in this figure, the $\al$ space
minimization is shown at the same density.
In Figure 2 we show the corresponding curve for the weakly localized
structure at the same density. The $\al$-space variation is shown as an
inset in this figure where each value corresponds to the vacancy concentration
obtained as a partial minimum in the $\cp$ space. In this case the optimum
value of $\cp$ is obtained at $0.946$, corresponding $\De f$ value being
$-1.325$.
The effect of vacancies on the stability of the weakly localized structure
is not as significant as that observed for the highly localized structures.
This is because
in the weakly localized structures the Gaussian profiles of the particles are
already too smeared. The vacancies are not expected to alter the density
structure to a large extent and hence the free energy values do not change
by a large percentage. Nevertheless, the existence of a minimum in the $\cp$
space at a value lesser than one indicates the thermodynamic preference
of the amorphous state for a structure characterized by a homogeneous
presence of vacancies and thus establishing the ``free-volume'' picture
of supercooled states from a density functional approach.

The two class of minima observed in amorphous structures represent two different
stationary states of the system and the dynamics of the system should show
different characteristics around these minima. This is indicated by the two
different type of behavior shown by the variation of vacancy concentration
with respect to the average density. It follows a linear relation with the
average
particle density of the highly localized amorphous structure. This is shown
in the Figure 3, where the solid line is the linear curve fitting to the
variation of $\cp$ versus $\r_o$ for the state corresponding to the high
$\al$ minimum. This predicts that the system will completely freeze, i.e.,
the free volume concentration will vanish, at packing fraction $\eta=.62$.
The Figure 4 illustrates the corresponding curve for the weakly localized
structure, corresponding to the lower width parameter values. The free volume
concentration shows a power law decrease with the 
increase in the density. The solid line of the curve depicts this functional
variation that shows the divergence at $\eta=.58$ with an exponent $.128$.

In Figure 5 we have shown the variation of the free energy versus the average
density for both the highly localized and the weakly localized amorphous
structures. The metastability of the highly localized amorphous structure
starts coming close to that of the weakly localized state at higher
densities due to the induction of free volumes in the structure. 
The highly localized state starts becoming more stable
than the homogenous state after density $\r_o=1.12$. In the similar density
structures considered without defects this density has been reported to be
$1.14$ \cite{SW}. The Figure 6 illustrates the variation of the $\al$ with
$\cp$ at density $\r_o=1.08$. This $\al$ corresponds to the free energy minima
at a given value of the defect concentration
$\cp$. The monotonic decrease depicts a preference for decreasing localization
in the system with decrease in defects concentration $\Dev$.

\section*{Discussion}
We have investigated various features  of the supercooled states by considering
the presence of a finite void concentration in an amorphous lattice. In
formulating the test density function  of the amorphous structure we have
basically extended the idea used by Jones and Mohanty \cite{UM}, who considered
defects in crystal structures.
By taking our cue from the Free Volume Theory, we have formulated
the corresponding theory in a Density Functional Formalism.
Important implications are drawn regarding  the structural aspects of
such states from this study.
The existence of the Free energy minima with respect to $\cp$ demonstrates the 
thermodynamic preference of the amorphous state for a particular
concentration of free volume corresponding to a specific density. That this brings
a decrease in the Free energy of the structure shows that the metastability
of the supercooled states is dependent upon the ``free volume'' present in
the structure. The weakly localized states which correspond to the free energy
minimum found in the region of small width parameter, although, do not 
gain stability by a significant percentage. This is reasonable since the
presence of a small concentration
of vacancies does not alter the density structure characterized by wide
Gaussians. Thus the free energy does not change much although a clear
minimum is observed with respect to the defect parameter $\cp$ at a non-zero
value of $\Dev$. At higher densities the stability of the highly localized
states comes closer to that of the other and thus demonstrating that at
higher densities, the supercooled state is likely to be characterized by the
more localized density structures.
We do not use here any of the more recent density functional theories like the
MWDA ( modified weighted density approximation ) \cite{DA}.
MWDA involves the {\em global} mapping of
the inhomogeneous structure to a homogenous structure satisfying certain
criteria in an effective medium type approximation.
Thus it is unsatisfactory for systems characterized by local
inhomogeneous effects like the presence of a small concentration of
vacancies as is done here.
Thus for qualitative investigation purpose, our use of RY functional approach
is well justified in this study.

We would also like to indicate here that the major criticism of the results
of the similar calculation done for the crystal structure \cite{UM}, was the
prediction of quite large values of $\Dev$. These were not in compliance with
the experimental findings which predicted the defect densities in the crystal
to be lesser than at least two orders of that predicted by Jones and Mohanty.
In the present study, although we have used the mathematical formulation of
the  density function, its physical interpretation is entirely different. The
defects in this study are a measure of the average amount of unoccupied volume
present in the amorphous structure. Its concentration is actually indicative
of the measure of dynamics possible in the system at a particular density.
The vanishing of this free volume concentration is interpreted as a complete
freezing of motion in the system and hence the relevance of this concept is
not wavered by the critical value of $\Dev$ obtained.

Since the free volume concentration is a measure of the transport
properties in a system, qualitative indications can also be drawn with respect
to the dynamical aspects. This study indicates that the decrease of the free
volume can in general be associated with increase in viscosity of the system.
The variation of free volume with density/temperature is useful in extracting
the temperature dependence of the transport coefficients. We observe two
different   type of behavior for the two different free energy minima.
The highly localized structures display a linear decrease in the free volume
concentration with the increase in density.
On the other hand, the weakly localized structures, characterized by the
low value of the width parameter, show a power-law decrease in the free volume
concentration with respect to density. It is well known that a large number
of laboratory systems and computer simulations of simple systems show a power
law divergence of viscosity with decrease in temperature.
It is also worth considering here that the dynamical studies done
using the Mode- Coupling Theories \cite{SPD3} also predict such a relation between
the the transport properties and the density for low viscosity hard sphere
systems.
Thus this result suggests that the low $\al$ Free energy minima may
correspond to the stationary state around which the fluctuations
are considered in the dynamical studies coming under the realm of
Mode-Coupling Theories. Although this would require further investigation
and a definite relation to be established between the free volume concentration
and the transport properties.

\newpage{}
\section*{Figure Captions}
\noindent
Fig 1:
Difference in Free Energy per particle ($\Delta f$)
 ( in units of $ \beta^{-1} $ ) vs. $ \cp $ 
in the large $\al$ regime ($ {\r_o}^*=1.12 $). In the inset the corresponding
minimization is shown in the $\al$ space.

\vspace*{.5cm}
\noindent
Fig 2 :
Difference in Free Energy per particle ($\Delta f$)
 ( in units of $ \beta^{-1} $ ) vs. $ \cp $ in the low $\al$ regime
 ($ {\r_o}^*=1.12 $). The inset shows the variation in the $\al$ space.

\vspace*{.5cm}
\noindent
Fig 3:
Free Volume concentration $(\Dev)$ vs. density ${\r_o}^*$ for the
highly localized amorphous structure. The solid line depicts the
linear fit. 

\vspace*{.5cm}
\noindent
Fig 4:
Free Volume concentration $(\Dev)$ vs. density ${\r_o}^*$ for the
weakly localized amorphous structure. The solid line depicts the
power-law fit.

\vspace*{.5cm}
\noindent
Fig 5:
Difference in Free Energy per particle ($\Delta f$)
 ( in units of $ \beta^{-1} $ ) vs. $ \r_o$. The solid line
corresponds to the small width parameter, $\al$, free energy minimum and the dashed line
corresponds to the highly localized structure.

\vspace*{.5cm}
\noindent
Fig 6:
Variation of minimum $\al$ with the corresponding $\cp$  at density
$\r_o=1.08$.

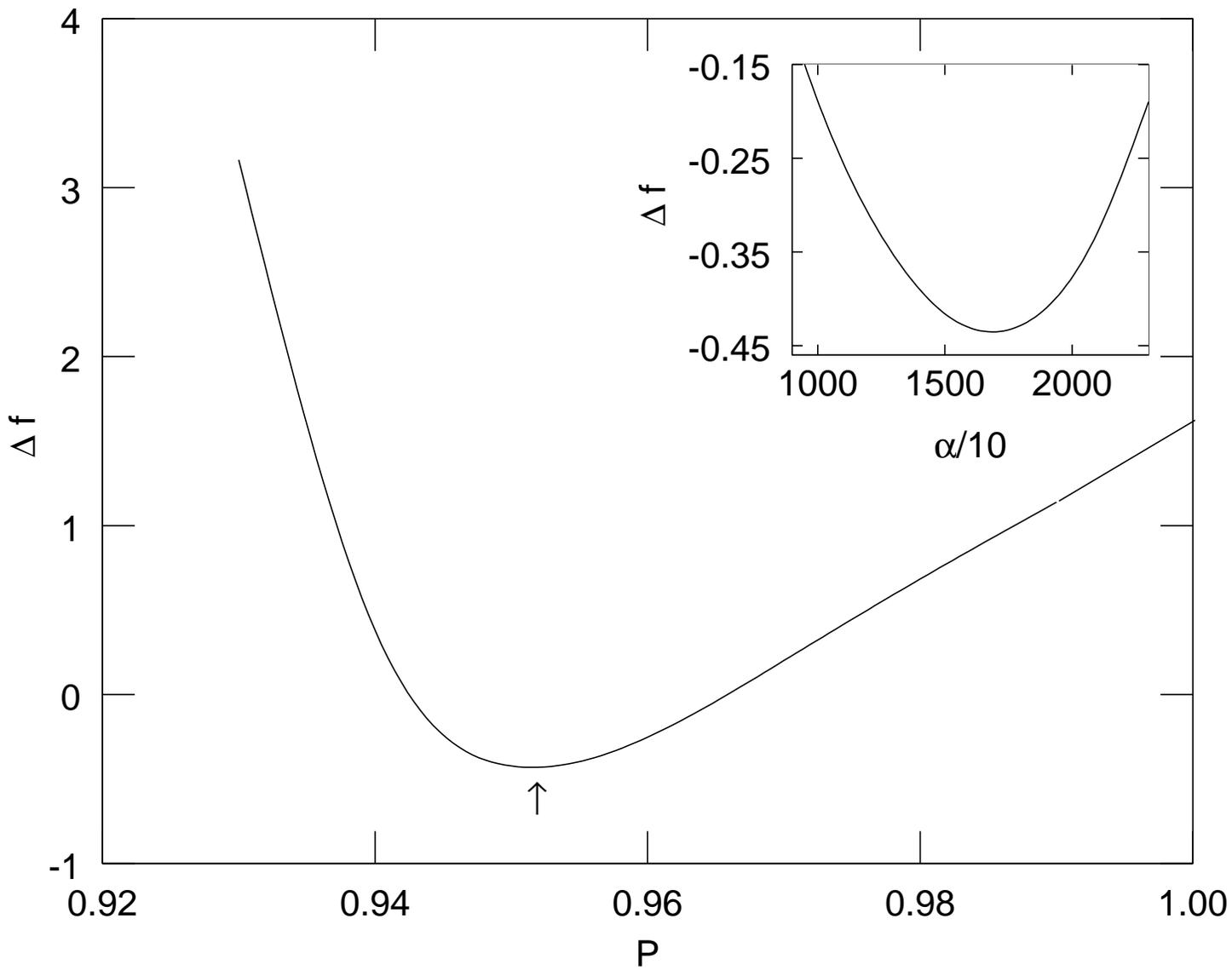
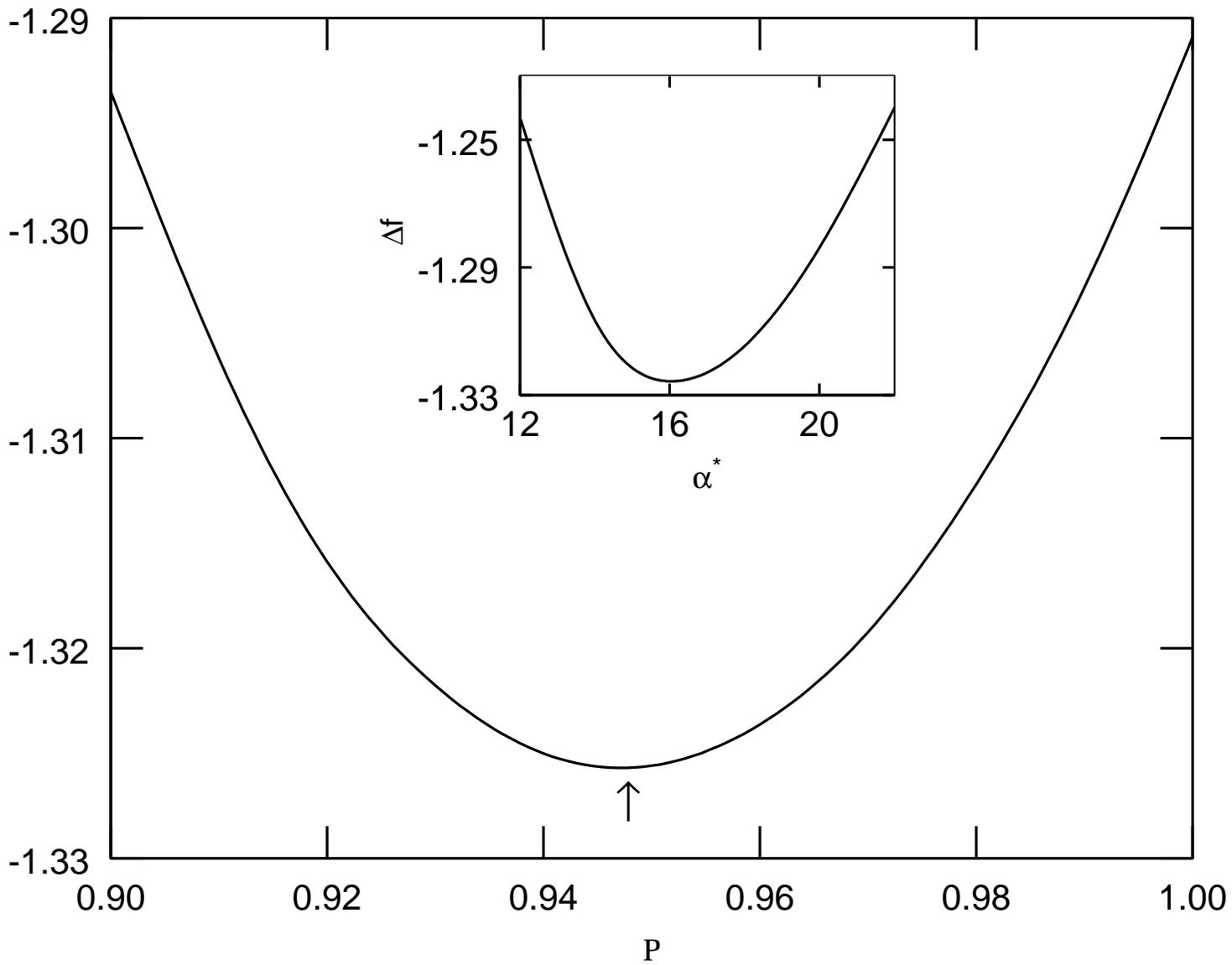
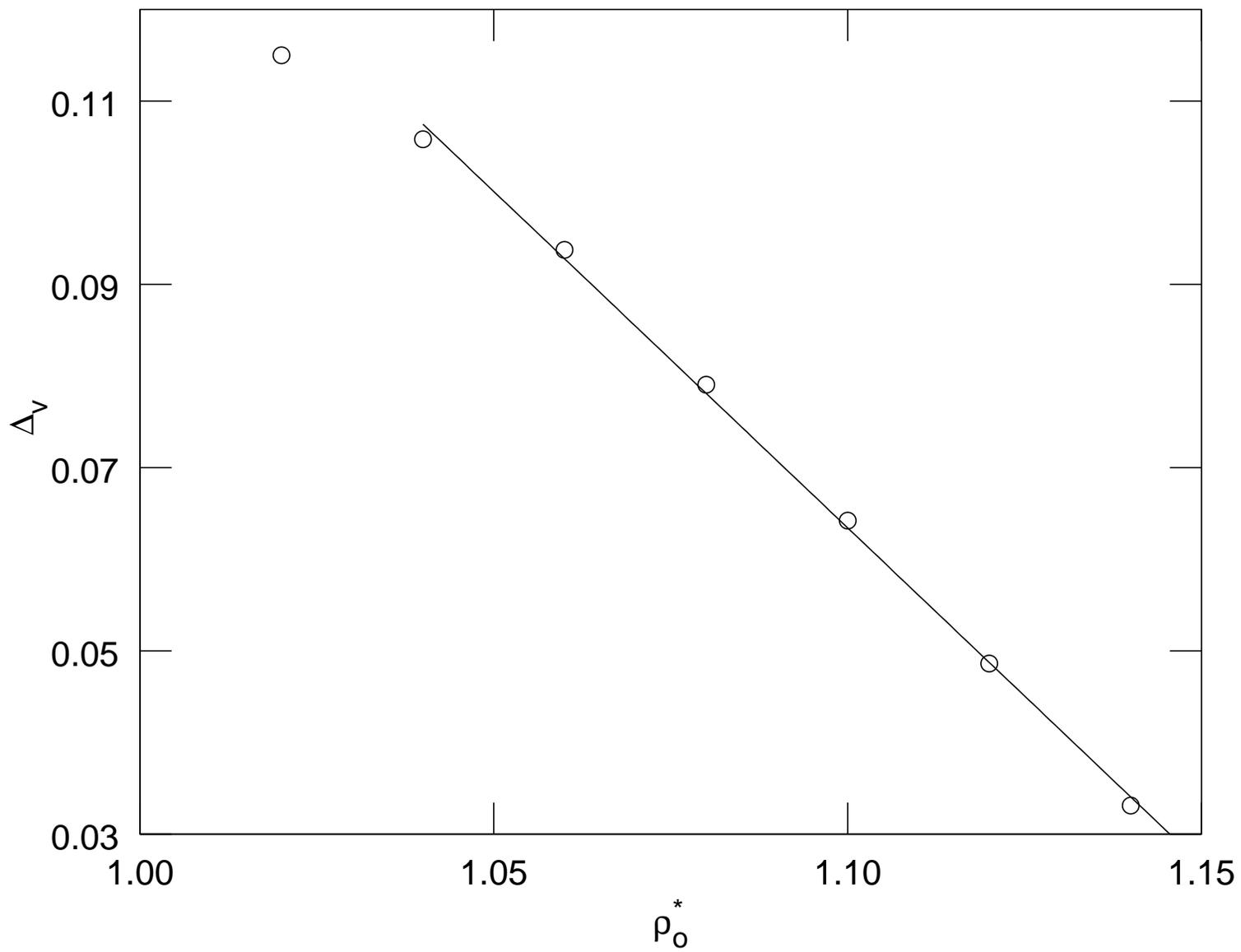
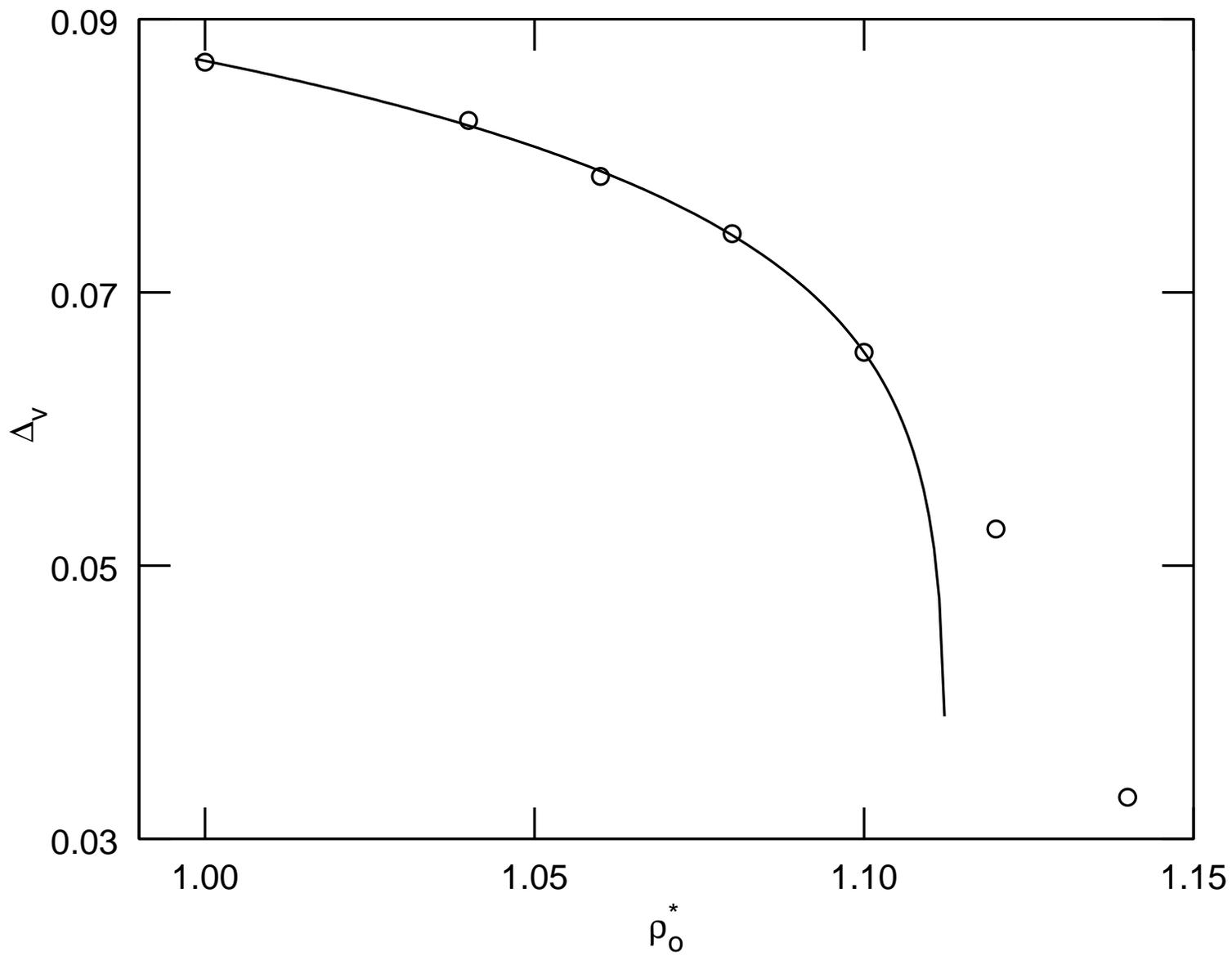
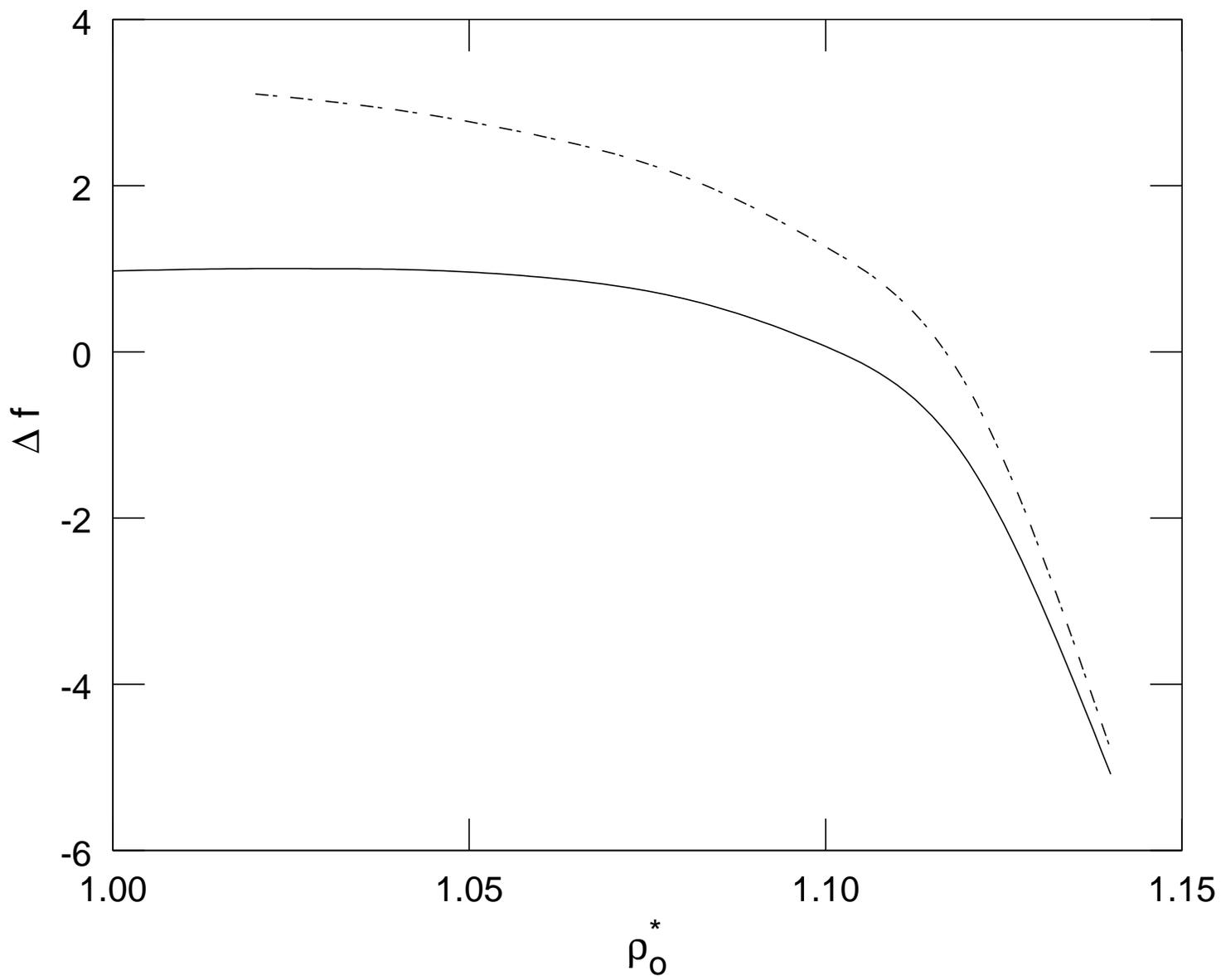
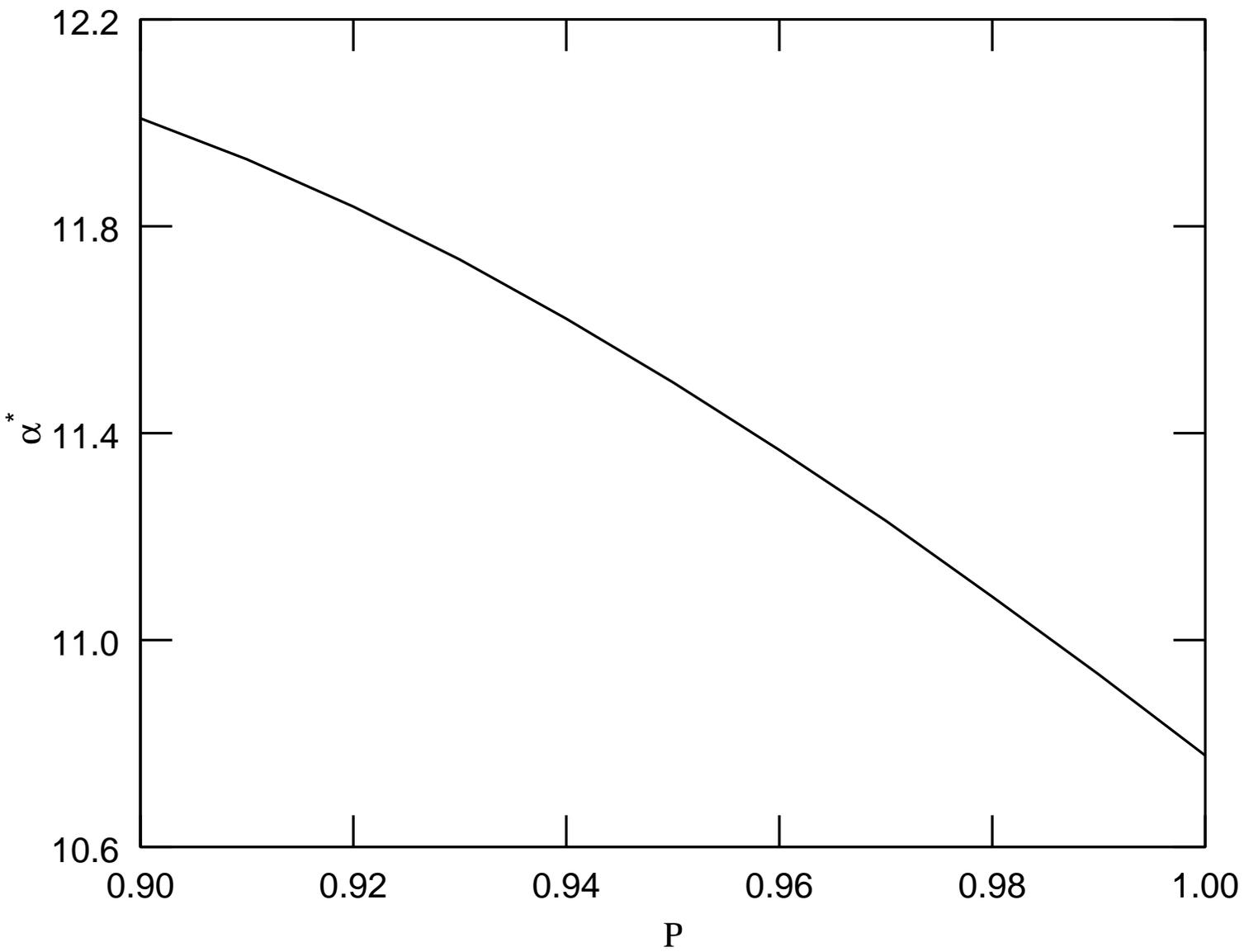
\end{document}